9 November 2025

# Enabling Frontier Lab Collaboration to Mitigate AI Safety Risks

Nicholas Felstead*

Frontier AI labs face intense commercial competitive pressure to develop increasingly powerful systems, raising the risk of a race to the bottom on safety. Voluntary coordination among labs — including by way of joint safety testing, information sharing, and resource pooling — could reduce catastrophic and existential risks. But the risk of antitrust scrutiny may deter such collaboration, even when it is demonstrably beneficial. This paper explores how U.S. antitrust policy can evolve to accommodate AI safety cooperation without abandoning core competition principles. After outlining the risks of unconstrained AI development and the benefits of lab–lab coordination, the paper analyses potential antitrust concerns, including output restrictions, market allocation, and information sharing. It then surveys a range of legislative and regulatory reforms that could provide legal clarity and safe harbours that will encourage responsible collaboration.

---

* BEnvs, JD (Melb); LLM (Columbia). AI Policy Fellow, Center for Law & AI Risk. Thanks to Professor Peter Salib, Ben Gantt, Ben Robinson, Zelda Hollings, and Tristan Lockwood for their insights. Views and errors are mine alone.

I    Introduction

Artificial intelligence (AI) offers tremendous promise to individuals and society. But without proper safeguards in place, it also presents the real possibility of catastrophe. AI companies are currently competing in a furious race to develop artificial general intelligence (AGI) and artificial super intelligence (ASI).[1] There is a risk that these hyper-competitive dynamics will lead frontier AI labs to run a race to the bottom, where safety is sacrificed in favour of the next best product or service. In the context of frontier AI development, race dynamics could lead to catastrophic disaster.

Despite growing recognition from governments, industry and civil society that accelerated AI development poses existential risk,[2] there are significant political, economic, and legal barriers to promoting safe and responsible AI. These race dynamics are one such barrier.[3] Given the distance that exists between frontier AI labs on the one hand, and governmental institutions (domestic and international) on the other, a key lever to ensure the safe and responsible development of AI may come from within the labs.[4] Specifically, if frontier AI labs work *together* to develop safer, responsible AI, we can avoid a race to the bottom that may otherwise leave our society in crisis.

---

[1] These concepts are often ill-defined, though Google's definitions are instructive: "AGI possesses human-like intelligence and can perform any intellectual task that a human can. It is capable of learning, reasoning, and adapting to new situations; ASI surpasses human intelligence and can potentially solve problems that are currently beyond the capabilities of humans. For instance, an ASI system could potentially design highly efficient energy systems or develop new medical treatments. However, ASI is still largely theoretical and remains a topic of debate and speculation". See Google Cloud, *What is artificial general intelligence (AGI)?*, https://cloud.google.com/discover/what-is-artificial-general-intelligence?hl=en.

[2] See generally Charlotte Stix and Matthijs M Maas, *Bridging the gap: the case for an 'Incompletely Theorized Agreement' on AI policy* (2021) 1 AI and Ethics 261; Thorin Bristow and Luke Thorburn, *Don't slip into binary thinking about AI* (2023); Emily M Bender and Alex Hanna, *The AI Con: How To Fight Big Tech's Hype and Create the Future We Want* (2025).

[3] See Amanda Askell et al, *The Role of Cooperation in Responsible AI Development* (2019) 9: "The race to the bottom on safety … is a collective action problem: a situation in which all agents would be better off if they could all cooperate with one another, but each agent believes it is in their interest to defect rather than cooperate." See also OpenAI's Charter: "We are concerned about late-stage AGI development becoming a competitive race without time for adequate safety precautions. Therefore, if a value-aligned, safety-conscious project comes close to building AGI before we do, we commit to stop competing with and start assisting this project.": OpenAI, *Charter*, https://openai.com/charter/.

But there are real concerns within frontier AI labs that collaborating with competing labs may raise antitrust scrutiny from domestic and global regulators. Anthropic recently advocated that "[r]egulators should issue guidance on permissible AI industry safety coordination given current antitrust laws. Clarifying how private companies can work together in the public interest without violating antitrust laws would mitigate legal uncertainty and advance shared goals".[5] Even if concerns about breaching antitrust laws are not well-founded, it remains the case that the mere potential for antitrust action may act as a deterrent for AI safety collaboration, and lead to dangerous race dynamics.

The U.S. Government should strike an appropriate balance between competition regulation and AI safety goals, and create specific mechanisms for frontier AI labs to coordinate and collaborate for the benefit of society at large without compromising the goals of competition policy. In particular, the U.S. Government could consider legislative solutions modelled on existing frameworks for information sharing and collaborative research and development (R&D), and regulatory guidance based on existing domestic tools.

---

[4] This lab-first approach is not designed to minimise the role that governmental bodies can play. Various provisions of the EU AI Act raise the floor on safety practices in a coordinated manner — for example, through the model evaluation, mitigation, and reporting requirements for providers of general-purpose AI models with systemic risk (art 55). I am grateful to colleagues that highlighted this collaboration "forcing function".

[5] Anthropic, Submission to the National Telecommunications and Information Administration (2024).

The purpose of this paper is to survey and draw attention to a range of possible proposals, with the aim of generating discussion and laying a foundation for further research. This paper proceeds as follows. Part II sets out the background and argues that AI safety collaboration is both desirable and necessary. Part III summarises foundational principles of competition policy and outlines the key instruments and bodies at issue. Part IV explores specific potential antitrust concerns raised by AI lab collaboration. Part V proposes potential legislative and regulatory changes to address these concerns. Part VI concludes.

## II    Why AI safety coordination is necessary

AI is already proving to be a transformative development in society. It is increasingly commonplace to say that "AI could accelerate scientific discoveries, revolutionize healthcare, enhance our education system, and create entirely new domains for human creativity and innovation".[6] It is similarly trite to note that these opportunities also come with significant challenges: the real threat of labour market disruptions raise serious questions about the future of work and the need to rethink societal expectations about the relationship between individuals and the State; ethical questions continue to arise, including those concerning bias in automated government decision-making, data rights, and AI-enabled mass surveillance; and there is a broad set of security, misuse and control challenges that rapid frontier model development pose to society.

---

[6]    Anthropic, *Announcing our updated Responsible Scaling Policy* (Oct. 15, 2024), https://www.anthropic.com/news/announcing-our-updated-responsible-scaling-policy.

This last category is often described as "catastrophic risk". It includes a model's vulnerability to exploitation, such as through prompt injection[7] or jailbreaking.[8] Relatedly, a model's capability to cause harm through facilitating cyberattacks[9] or contributing to the development or deployment of chemical, biological, radiological or nuclear weapons.[10] A particularly acute concern involves loss of human oversight,[11] where models engage in deceptive or manipulative behavior such as faking alignment,[12] or exhibit emergent opacity in reasoning and chain-of-thought processes,[13] or unintended autonomous self-replication.[14] Additionally, frontier models may significantly amplify downstream risk by enabling the development of even more dangerous systems. Finally, these models may be susceptible to unauthorised access, including by state and non-state actors.[15]

---

[7] Matt Sutton and Damian Ruck, *Indirect Prompt Injection: Generative AI's Greatest Security Flaw* (2024), https://cetas.turing.ac.uk/publications/indirect-prompt-injection-generative-ais-greatest-security-flaw.

[8] Microsoft Threat Intelligence, *AI jailbreaks; What they are and how they can be mitigated* (2024), https://www.microsoft.com/en-us/security/blog/2024/06/04/ai-jailbreaks-what-they-are-and-how-they-can-be-mitigated/.

[9] See Dan Hendrycks et al, *An Overview of Catastrophic AI Risks* (Center for AI Safety) 14.

[10] See David Luckey et al, *Mitigating Risks at the Intersection of Artificial Intelligence and Chemical and Biological Weapons* (2025); Elliot McKernon, *AI and Chemical, Biological, Radiological, & Nuclear Hazards* (2024), https://www.convergenceanalysis.org/ai-regulatory-landscape/ai-and-chemical-biological-radiological-and-nuclear-hazards.

[11] See generally Dan Hendrycks et al, *An Overview of Catastrophic AI Risks* (Center for AI Safety) 34.

[12] Ryan Greenblatt et al, *Alignment faking in large language models* (2024), https://arxiv.org/pdf/2412.14093.

[13] Fabien Roger and Ryan Greenblatt, *Preventing Language Models from Hiding Their Reasoning* (2023), https://arxiv.org/pdf/2310.18512; Tamera Lanham et al, *Measuring Faithfulness in Chain-of-Though Reasoning* (2023), https://storage.prod.researchhub.com/uploads/papers/2023/08/01/2307.13702.pdf.

Collectively, these risks underscore the urgent need for coordinated safety measures at the frontier of AI capability.

A      *Race dynamics risk safe and responsible AI*

There is a real imperative for rigorous safety research to ensure transparency, accountability, and reliability of AI systems. Using OpenAI and Anthropic as examples, each company has made a public commitment to safe and responsible AI development.

OpenAI's charter states that:[16]

> *OpenAI's mission is to ensure that artificial general intelligence (AGI)—by which we mean highly autonomous systems that outperform humans at most economically valuable work—benefits all of humanity. We will attempt to directly*

---

[14] See generally Caroline Delbert, *A.I. Can Now Replicate Itself without Human Help. Can We Control What Comes Next?* (2025), https://www.popularmechanics.com/science/a64288856/ai-replication/.

[15] See Sarah Kreps, *Democratizing Harm: Artificial Intelligence in the Hands of Nonstate Actors* (2021), https://www.brookings.edu/wp-content/uploads/2021/11/FP_20211122_ai_nonstate_actors_kreps.pdf; Valerie Wirtschafter, *The implication of the AI boom for nonstate armed actors* (2024), https://www.brookings.edu/articles/the-implications-of-the-ai-boom-for-nonstate-armed-actors/; Tom Davidson et al, *AI-Enabled Coups: How a Small Group Could Use AI to Seize Power* (2025), https://www.forethought.org/research/ai-enabled-coups-how-a-small-group-could-use-ai-to-seize-power.pdf.

[16] OpenAI, *Charter*, https://openai.com/charter/.

> *build safe and beneficial AGI, but will also consider our mission fulfilled if our work aids others to achieve this outcome.*

Anthropic's self-proclaimed purpose is:[17]

> *We believe AI will have a vast impact on the world. Anthropic is dedicated to building systems that people can rely on and generating research about the opportunities and risks of AI.*

Other leading companies have made similar public declarations.[18] Notwithstanding these public statements, the reality is that these are private companies with an incentive to beat out competitors and gain market share. Frontier AI labs boast multi-billion dollar valuations and attract leading talent with ever-growing pay packages.[19] While there may be some profit motive to develop safe AI (if the market or fear of costly regulatory action demands it), fierce competition and race dynamics may distort the cost–benefit ratio of responsible AI development.[20]

---

[17] Anthropic, *Our Purpose*, https://www.anthropic.com/company.

[18] Microsoft: "We're committed to developing AI systems in a way that is transparent, reliable, and worthy of trust": Microsoft, *Principles and approach*, https://www.microsoft.com/en-us/ai/principles-and-approach. Deepmind: "Guided by the scientific method and with a holistic approach to responsibility and safety, we're working to ensure AI benefits everyone and helps solve the biggest challenges facing humanity": Google DeepMind, *Build AI responsibly to benefit humanity*, https://deepmind.google/about/.

[19] See Mark Gurman and Riley Griffin, *Meta Poached Apple's Ruoming Pang with Pay Package over $200 Million*, Bloomberg (July 9, 2025), https://www.bloomberg.com/news/articles/2025-07-09/meta-poached-apple-s-pang-with-pay-package-over-200-million?embedded-checkout=true.

[20] See Amanda Askell et al, *The Role of Cooperation in Responsible AI Development* (2019) 8.

So whether deliberate or not, there is a real risk that competitive pressures will drive frontier AI labs to engage in a race to the bottom in their research and development.[21] OpenAI recognises in its Charter that it is "concerned about late-stage AGI development becoming a competitive race without time for adequate safety precautions".[22] In short, the risk is that model development outpaces safeguards, safety investments, and ethical baselines. Sensible guardrails can help mitigate those risks and cool the race dynamics between frontier AI labs. One way to avoid this race to the bottom would be for AI labs to coordinate and collaborate on safety research.

B  *Why collaborate?*

Research collaboration between competitors can have real benefits. The European Union describe the benefits in these succinct terms:[23]

> *Cooperation in joint or paid-for research and development and in the exploitation of the results is most likely to promote technical and economic progress if the parties contribute complementary skills, assets or activities to the cooperation.*
>
> *Consumers can generally be expected to benefit from the increased volume and effectiveness of research and development through the introduction of new or improved products,*

---

[21] See generally Wim Naudé and Nicola Dimitri, *The race for an artificial general intelligence: implications for public policy*, 35 AI & Society 367 (2020).
[22] OpenAI, *Charter*, https://openai.com/charter/.
[23] Commission Regulation 2023/1066 (2023), [7]–[8].

*technologies or processes, a quicker launch of such products, technologies or processes, or a reduction of prices brought about by new or improved products, technologies or processes.*

Until 2024, the joint position of the Department of Justice (DOJ) and the Federal Trade Commission (FTC) was that:[24]

*In order to compete in modern markets, competitors sometimes need to collaborate. Competitive forces are driving firms toward complex collaborations to achieve goals such as expanding into foreign markets, funding expensive innovation efforts, and lowering production and other costs.*

*…*

*The Agencies recognize that consumers may benefit from competitor collaborations in a variety of ways. For example, a competitor collaboration may enable participants to offer goods or services that are cheaper, more valuable to consumers, or brought to market faster than would be possible absent the collaboration. A collaboration may allow its participants to better use existing assets, or may provide incentives for them to make output-enhancing investments that would not occur absent the collaboration.*

Collaboration between frontier AI labs can similarly promote progress and mitigate catastrophic and existential risks. There is much to be gained from a coordinated approach to safety, which can lead to greater risk reduction than incremental improvements at individual labs. Enmity "can increase the danger of an AI-disaster", and collaboration can reduce hostilities between AI labs and produce better outcomes.[25] Collaboration can also foster mutual interdependence, where each lab becomes invested in the others' commitment to safety, creating shared incentives to uphold high standards and reduce the likelihood of reckless or unilateral actions that could jeopardise collective progress. This is not to suggest that collaboration would result in homogeneous product and service offerings — individual labs would still be incentivised to lean into their comparative advantage and compete with each other, but with a joint (and practical) safety-by-design approach to R&D.

---

[24] Federal Trade Commission and Department of Justice, *Antitrust Guidelines for Collaborations Among Competitors* (2000) ("Guidelines"). These Guidelines are discussed in detail below at page **XX**.

### C     *Forms of collaboration*

There are several ways in which frontier AI labs can collaborate. A prominent example is through a third party intermediary, such as the Frontier Model Forum (FMF). The FMF is an "industry non-profit dedicated to advancing the safe development and deployment of frontier AI systems".[26] Members collaborate through the FMF by way of initiatives including an "AI Safety Fund" which supports independent research to promote responsible development and minimise risks of frontier AI models, and through a voluntary "inter-firm information-sharing" arrangement.[27] This paper focuses on direct lab–lab collaboration. An initiative like the FMF is laudable and can facilitate coordination concerning high-level principles, joint policy communication, and some information sharing.[28] Direct lab–lab relationships can facilitate deep collaborations, binding commitments, and a hands-on exchange of technical expertise and resources. The proposals set out in this paper can complement the role of a convening body like the FMF, as legislative and regulatory guidance can also promote intermediary coordination.[29]

---

[25] Stuart Armstrong, Nick Bostrom and Carl Shulman, *Racing to the Precipice: A Model of Artificial Intelligence Development*, Future of Humanity Institute (2013).
[26] The FMF members are Amazon, Anthropic, Google, Meta, Microsoft, and OpenAI: Frontier Model Forum, *Membership*, https://www.frontiermodelforum.org/membership/.

Several forms of lab–lab collaboration below are described in general terms. Further work is necessary to set out the details of how any formal agreement would be structured.

| Collaboration | Explanation | Rationale |
|---|---|---|
| Safety testing / cross-red teaming[30] | Joint development and execution of standardised AI evaluations. | Improves detection of dangerous behavior; ensures consistency; avoids blind spots unique to individual labs. |

---

[27] There are very few public details about the mechanisms of information sharing between FMF members: Frontier Model Forum, *FMF Announces First-Of-Its-Kind Information-Sharing Agreement* (Mar. 28, 2025), https://www.frontiermodelforum.org/updates/fmf-announces-first-of-its-kind-information-sharing-agreement/.

[28] Frontier Model Forum, *FMF Information-Sharing*, https://www.frontiermodelforum.org/information-sharing/.

[29] See, eg, below at Part IV.A.

[30] See, eg, OpenAI, *Findings from a pilot Anthropic–OpenAI alignment evaluation exercise: OpenAI Safety Tests* (2025), https://openai.com/index/openai-anthropic-safety-evaluation/.

| | | |
|---|---|---|
| Incident sharing | Disclosure of safety incidents and near-misses between labs. | Accelerates learning from failures; prevents repeated mistakes; promotes transparency and early issue detection. |
| Information sharing | Exchange of alignment methods, evaluation metrics, and best practices. | Facilitates rapid adoption of effective safety techniques; avoids redundant work; raises the safety floor across the industry. |
| Compute / resource pooling | Shared access to compute infrastructure for safety-related work. | Enables intensive and faster safety testing; supports smaller labs; prevents safety from being deprioritised due to resource constraints. |
| Joint model development | Collaborative creation of models for safety evaluation and benchmarking. | Builds shared evaluation tools; enhances cross-lab comparisons; supports the development of safety-specific testing models. |
| Developmental pause[31] | Agreement to pause development if certain safety thresholds are breached. | Provides time for investigation and mitigation; avoids escalation; signals responsible development norms across competing labs. |

---

[31] See Jide Alaga and Jonas Schuett, *Coordinated pausing: An evaluation-based coordination scheme for frontier developers* (2023) ("*Coordinated pausing*").

| Standard-setting | Development and adoption of open technical standards for safety evaluations and model governance. | Raises the safety baseline; enables comparability and auditability; accelerates shared learning. |

In 2025, OpenAI and Anthropic collaborated on a joint evaluation, where each company ran its own internal evaluations on the other company's public models.[32] One goal of this "first-of-its-kind" collaboration was to demonstrate how frontier AI labs can work together on safety and alignment. As OpenAI noted, the exercise "supports accountable and transparent evaluation, helping to ensure that each lab's models continue to be tested against new and challenging scenarios".[33] One potential shortfall of this collaboration is that the evaluations were on each other's *publicly* released models. The proposals introduced in this paper may lay the foundation for deeper collaboration that can allow for more proactive monitoring and evaluations.

---

[32] OpenAI, *Findings from a pilot Anthropic–OpenAI alignment evaluation exercise: OpenAI Safety Tests* (2025), https://openai.com/index/openai-anthropic-safety-evaluation/; Anthropic, *Findings from a Pilot Anthropic–OpenAI Alignment Evaluation Exercise* (2025), https://alignment.anthropic.com/2025/openai-findings/.
[33] OpenAI, *Findings from a pilot Anthropic–OpenAI alignment evaluation exercise: OpenAI Safety Tests* (2025), https://openai.com/index/openai-anthropic-safety-evaluation/.

**III    Antitrust foundations**

Competition regulation is an enduring endeavour that creates a framework within which our economy and society function.[34] It is designed to protect the competitive process,[35] which should lead to lower prices, greater consumer choice, improved working conditions and wages, and improved product and service quality. Competition policy helps "protect competition as the most appropriate means of ensuring the efficient allocation of resources — and thus efficient market outcomes — in free market economies".[36]

The relationship between AI and competition policy is complex. On the one hand, competition policy can help drive innovation and avoid undesirable consolidations of power in a few large companies.[37] And the AI industry already suffers from a concentration problem[38] — for example, TSMC enjoys a "near monopoly" over advanced semiconductors,[39] ASML wields a "near-total monopoly" over extreme ultraviolet lithography (an essential input in chip manufacturing),[40] the DOJ is investigating Nvidia for alleged monopolistic practices,[41] and RAND considers that the market for foundation models exhibits relevant characteristics of a natural monopoly.[42]

---

[34] *See* Michael Gvozdenovic & Stephen Puttick, *Perspectives on Part IV*, in CURRENT ISSUES IN COMPETITION LAW: PRACTICE AND PERSPECTIVES 3 (Michael Gvozdenovic & Stephen Puttick, eds., 2021); ARIE FREIBERG, REGULATION IN AUSTRALIA 42 (2017).

[35] See, eg, *Brown Shoe Co. v. United States*, 370 U.S. 294, 320 (1962); *Australian Competition and Consumer Commission v Pacific National Pty Ltd (No 2)* [2019] FCA 669, [1261].

[36] OECD, *Interim Report on Convergence of Competition Policies* (1994) 8–9.

[37] See Jack Corrigan, *Promoting AI Innovation Through Competition: A Guide to Managing Market Power* (2025). See generally Anu Bradford, *The False Choice between Regulation and Innovation* (2024).

[38] These examples, among others, will be explored in a forthcoming piece: Nicholas Felstead, *AI's Competition Problem: Here We Go Again* (2025).

[39] See Debby Wu, *TSMC's Dominance Is Starting to Worry More Than Just Rivals* (2024), https://www.bloomberg.com/news/newsletters/2024-10-21/tsmc-s-dominance-is-starting-to-worry-more-than-just-rivals.

On the other hand, and notwithstanding the benefits of joint safety research, the threat of being exposed to regulatory antitrust scrutiny looms large and may deter frontier AI labs from exploring avenues to collaborate. As the DOJ and FTC once recognised, "a perception that antitrust laws are skeptical about agreements among actual or potential competitors may deter the development of procompetitive collaborations".[43] Similarly, Anthropic recently noted that "[c]larity on antitrust regulation would help determine whether and how AI labs can collaborate on safety standards".[44]

Understanding the basics of U.S. antitrust law is foundational for the balance of this paper, and for frontier AI labs considering collaboration. These laws aim to maintain competitive markets by preventing practices that harm consumers through higher prices, reduced quality, or limited choice.

---

[40] See Julian West, *ASML Holding: The Indispensable Monopoly Powering the AI Revolution* (2025), https://www.ainvest.com/news/asml-holding-indispensable-monopoly-powering-ai-revolution-2507/#; Renato Neves, *ASML: The EUV Lithography Giant Navigating Challenges* (2024), https://finance.yahoo.com/news/asml-euv-lithography-giant-navigating-130136031.html.

[41] Associated Press, *Nvidia is facing an antitrust probe from US regulators amid competitor complaints, report says* (Aug. 2, 2024); Anissa Gardizy, Stephanie Palazzolo, and Amir Efrati, *Nvidia Faces DOJ Antitrust Probe Over Complaints from Rivals* (Aug. 1, 2024), https://www.theinformation.com/articles/nvidia-faces-doj-antitrust-probe-over-complaints-from-rivals

[42] RAND, *Evaluating Natural Monopoly Conditions in the AI Foundation Model Market* (2024).

[43] Department of Justice and Federal Trade Commission, *Guidelines* at 1.

[44] Anthropic, Submission to the National Telecommunications and Information Administration (2024).

As Francis and Sprigman note, the "core of the U.S. antitrust system … consists of three main prohibitions: a rule against anticompetitive agreements (agreements in "restraint of trade") in Section 1 of the Sherman Act; a rule against the improper creation or maintenance of monopoly power ("monopolization") in Section 2 of the Sherman Act; and a rule against anticompetitive mergers and acquisitions in Section 7 of the Clayton Act. We will call these the three pillars of U.S. antitrust".[45] For the purposes of competitor collaborations, the first pillar — a prohibition against agreements in restraint of trade — is of particular importance. Section 1 of the Sherman Antitrust Act relevantly provides as follows:

> *Every contract, combination in the form of trust or otherwise, or conspiracy, in restraint of trade or commerce among the several States, or with foreign nations, is declared to be illegal.*

A court faced with a §1 challenge involving an agreement amongst competitors will apply one of two standards of review.[46] A court will usually consider bare or nakedly

---

[45] Daniel Francis and Christopher Jon Sprigman, *Antitrust: Principles, Cases, and Materials* (2024) 2. Section 5 of the FTC Act also provides that "Unfair methods of competition in or affecting commerce, and unfair or deceptive acts or practices in or affecting commerce, are hereby declared unlawful": 15 U.S.C. §45(a)(1).
[46] For completeness, courts can also engage in "quick look" review in cases where "an observer with even a rudimentary understanding of economics could conclude that the arrangements in question would have an anticompetitive effect on customers and markets": *California Dental Association v. Federal Trade*

anticompetitive agreements as "per se" illegal, without regard to purpose or effect.[47] Agreements to divide markets, fix prices or restrict output will typically be considered under the per se standard. Other agreements will typically be considered under the "rule of reason", which involves the court undertaking an evaluative inquiry to determine whether any procompetitive effects of the agreement outweigh the anticompetitive effects.[48] Most antitrust cases in the U.S. are analysed under the rule of reason approach.[49]

The antitrust enforcement architecture is straightforward. The DOJ and FTC each have jurisdiction over civil antitrust cases (only the DOJ can pursue criminal cases). State Attorneys-General can bring federal antitrust cases on behalf of their residents by way of a *parens patriae* suit, or enforce their own state antitrust law. Finally, private plaintiffs can also bring federal and state antitrust cases.[50]

---

*Commission*, 526 U.S. 756, 770 (1999). See also Jonathan Berman, *The Failure of "Quick Look" Analyses of Antitrust Claims*, 14 American University Business Law Review 79.

[47] See the Sixth Circuit's comment in *In re Cardizem CD Antitrust Litigation*, 332 F.3d 896, 909 (6th Cir. 2003): "[T]he virtue/vice of the per se rule is that it allows courts to presume that certain behaviors as a class are anticompetitive without expending judicial resources to evaluate the actual anticompetitive effects or procompetitive justifications in a particular case".

[48] Rule of reason analysis requires a "fact-intensive inquiry into the purpose and effect of the collaboration, including any business justifications": Federal Trade Commission, *Dealings with Competitors*, https://www.ftc.gov/advice-guidance/competition-guidance/guide-antitrust-laws/dealings-competitors.

[49] Daniel Francis and Christopher Jon Sprigman, *Antitrust: Principles, Cases, and Materials* (2024) 8.

[50] "In fact, most antitrust suits are brought by businesses and individuals seeking damages for violations of the Sherman or Clayton Act": Federal Trade Commission, *The Enforcers*, https://www.ftc.gov/advice-guidance/competition-guidance/guide-antitrust-laws/enforcers.

This background to U.S. antitrust law should provide an essential foundation to explore why frontier AI labs may be unwilling to collaborate, and to propose sensible and practical solutions.

## IV     Specific concerns raised by AI safety research collaboration

While AI safety research collaboration could bring significant societal benefits and reduce high-risk race dynamics, coordination between competitors may raise antitrust concerns. Collaboration agreements can harm competition and consumers by increasing participants' ability (and incentive) to raise prices above competitive levels, restrict output, reduce quality, or stifle innovation. These undesirable anti-competitive effects may come about from a variety of agreements and collaborations. Collaboration agreements could "limit independent decision making or combine the control of or financial interests in production, key assets, or decisions regarding price, output, or other competitively sensitive variables, or may otherwise reduce the participants' ability or incentive to compete independently".[51] Output restrictions, price fixing, market allocation, and information-sharing are some forms of conduct between competitors that may raise antitrust concerns. While each will be discussed in turn, output restrictions are the most interesting[52] and will be considered in more detail than other forms of conduct.

---

[51] Federal Trade Commission and Department of Justice, *Antitrust Guidelines for Collaborations Among Competitors* (2000).

[52] In my view.

## A  *Output restrictions*

Output restrictions involve agreements between competitors to limit the production, development, or availability of products or services. Output restrictions are a "paradigmatic example[] of [a restraint] of trade that the Sherman Act was intended to prohibit".[53] These are "ordinarily condemned … under an illegal *per se* approach, because the probability that these practices are anticompetitive is so high".[54] In some cases though, courts may decline to apply the *per se* rule in circumstances where they have "a lack of judicial experience with th[e] type of arrangement, [or where] a case involves an industry in which horizontal restraints on competition are essential if the product is to be available at all".[55]

In the context of AI safety collaborations, a particularly relevant course of conduct that may be construed as an output restriction would be a "developmental pause" or a

---

[53] NCAA v. Board of Regents, 468 U.S. 85, 107–8. See also Phillip Areeda, Louis Kaplow, and Aaron Edlin, *Antitrust Analysis: Problems, Text, and Cases* (2013) at 395; Daniel Francis and Christopher Jon Sprigman, *Antitrust: Principles, Cases, and Materials* (2024) at 169.

[54] NCAA v. Board of Regents, 468 U.S. 85, 100.

[55] NCAA v. Board of Regents, 468 U.S. 85, 100–101. As Justice Stevens observed:

> What the NCAA and its member institutions market in this case is competition itself — contests between competing institutions. Of course, this would be completely ineffective if there were no rules on which the competitors agreed to create and define the competition to be marketed. A myriad of rules affecting such matters as the size of the field, the number of players on a team, and the extent to which physical violence is to be encouraged or proscribed, all must be agreed upon, and all restrain the manner in which institutions compete.

Judge Easterbrook similarly considered that when "cooperation contributes to productivity through integration of efforts, the Rule of Reason is the norm": Polk Bros., Inc., v. Forest City Enterprises, Inc., 776 F.2d 185 (7th Cir. 1985).

similar agreement to delay model releases or limit capabilities in a coordinated manner. Alaga and Schuett considered the possibility of coordinated pausing in a paper proposing an "evaluation-based coordination scheme" for frontier AI labs.[56] In brief, that scheme includes the following steps:

- **Step 1: Dangerous capabilities evaluations**. Frontier AI models are evaluated for dangerous capabilities;

- **Step 2: Individual pausing**. Whenever, and each time, a model fails a set of evaluations, the developer pauses any further training and fine-tuning of that model. They also pause the development and deployment of similar models and do not publish related research;

- **Step 3: Coordinated pausing**. Other developers are notified whenever a model with dangerous capabilities has been discovered. They also pause the development and deployment of similar models and do not publish related research;

- **Step 4: Investigation during pausing**. The discovered capabilities are analyzed and adequate safety precautions are put in place;

- **Step 5: Resuming paused activities**. Developers only resume their paused activities if certain safety thresholds are reached.

---

[56] Alaga and Schuett, *Coordinated pausing*.

The proposed coordinated pause may raise antitrust concerns. Alaga and Schuett discuss four different versions of the coordination scheme, each of which would alter the potential level of antitrust scrutiny.

The first version is "*voluntary pausing*", where frontier AI labs make no commitments to pause and there are no legal requirements to pause (but would "face public pressure to do so").[57] Voluntary pausing raises little antitrust concern. A conspiracy or agreement is "the key element" of a §1 Sherman Act case.[58] If an individual frontier AI lab decided to pause development on a truly voluntary basis, in the absence of any express or implied agreement, it is unlikely to meet the statutory threshold of a "contract, combination …, or conspiracy".[59]

The second version is a "*pausing agreement*". In this version, frontier AI labs would negotiate a contract and "commit to commission a third party to run dangerous capabilities evaluations, notify the other contracting parties if a model fails a set of evaluations, and pause certain research and development activities until certain safety thresholds are reached."[60] There is a considerable risk with a pausing agreement. Because the parties would be agreeing to curtail research, development, and supply of models, the arrangement risks being deemed a prima facie horizontal output restriction — conduct that ordinarily draws *per se* condemnation under §1.

---

[57] Alaga and Schuett, *Coordinated pausing* at 7.
[58] Department of Justice, *Antitrust Resource Manual*: "The conspiracy or agreement to fix prices, rig bids or allocate markets is the key element of a Sherman Act criminal case. In effect, the conspiracy must comprise an agreement, understanding or meeting of the minds between at least two competitors or potential competitors, for the purpose or with the effect of unreasonably restraining trade. The agreement itself is what constitutes the offense; overt acts in furtherance of the conspiracy are not essential elements of the offense and need not be pleaded or proven in a Sherman Act case."
[59] This is especially so in circumstances where courts have rejected the view that tacit coordination and oligopolistic interdependence amounts to a §1 violation. See *Bell Atlantic Corp. v. Twombly*, 550 U.S. 544 (2007).

The third version involves a "*mutual auditor*" who evaluates frontier AI labs' models. Labs would make individual agreements with a mutual external auditor, who would "run dangerous capabilities evaluations on all frontier models they develop".[61] The developers would "commit to pause certain research and development activities if the auditor informs them that one of their models has failed a set of evaluations. … Inversely, they commit to pause certain research and development activities if the auditor notifies them that a model from another developer has failed a set of evaluations."[62] This cross-commitment may function like a coordinated output restraint, because one firm's conduct (in this case, evaluation failure) would suppress another firm's output. Demonstrating that an independent auditor and objective, non-discretionary evaluations could go some way to developing a colourable argument for rule of reason treatment, by analogy to contexts where horizontal coordination is necessary to product viability.[63]

---

[60] Alaga and Schuett, *Coordinated pausing* at 9.
[61] Alaga and Schuett, *Coordinated pausing* at 10.
[62] Alaga and Schuett, *Coordinated pausing* at 10.
[63] A topic for further exploration might involve consideration of whether a mutual auditor could serve as a standard-setting authority. Rule of reason analysis applies to standards development organizations engaged in standards development activities, pursuant to the National Cooperative Research and Production Act of 1993 (as amended by the Standards Development Organization Act of 2004).

The fourth version involves "*pausing requirements*" by way of new laws or regulations. It should suffice to note that frontier AI labs pausing development (and thereby restricting output) in compliance with a legal or regulatory requirement to do so would *not* violate antitrust law.

Although pausing development in agreement with, or in response to, a competitor's advancements may seem unusual for profit-driven companies, it is not a mere hypothetical. OpenAI commits in its Charter to a form of voluntary pausing, by way of its "Assist Clause":

> *[I]f a value-aligned, safety-conscious project comes close to building AGI before we do, we commit to stop competing with and start assisting this project. We will work out specifics in case-by-case agreements, but a typical triggering condition might be "a better-than-even chance of success in the next two years."*[64]

If the Assist Clause only contained a commitment to "stop competing with the project", then it may be construed as a purely unilateral action (similar to the abovementioned voluntary pause model). That unilateral action may have the effect of an anticompetitive output restriction, but it lacks the key element of an agreement. But OpenAI goes further and commits to "stop competing with *and start assisting* this project". Presumably, the competitor would have to consent (in some capacity) to OpenAI's assistance, given the clarification that OpenAI will "work out specifics in case-by-case agreements". Once operationalised, the Assist Clause might be considered an anticompetitive agreement to restrict output by limiting the production and release of a product, in concert with a competitor, contrary to §1 of the Sherman Act. Of course, there may be reason to believe that the Assist Clause is never enlivened. The Charter does not make clear what would constitute a "value-aligned, safety-conscious project", or when such an undefined project comes "close to building AGI". The Charter provides only that a "typical triggering condition *might* be 'a better-than-even chance of success in the next two years'".[65] In the absence of ex ante criteria, potential concerns about OpenAI shifting goalposts should be secondary to the threshold problem that no goalposts exist. At this point, it will have to suffice to say that the Assist Clause *could* raise antitrust issues, although the details of any antitrust action would turn on more specific facts.[66]

---

[64] OpenAI, *Charter*, https://openai.com/charter/.

B     *Price fixing*

Price fixing refers to agreements between competitors to directly or indirectly set prices or constrain pricing decisions. It is the paradigmatic form of conduct considered *per se*

---

[65] OpenAI, *Charter*, https://openai.com/charter/ (emphasis added). Some obvious questions arise: Will this "typical" trigger condition be implemented in actual agreements with competitors? At what point will OpenAI enter into such agreements? Has OpenAI already entered into any agreements? How will the parties determine that there is a "better-than-even" chance of success?

[66] A thorough examination of the Assist Clause, including a comparative analysis under US, EU, and Australian antitrust law, will be the subject of a forthcoming paper.

illegal under U.S. antitrust law,[67] regardless of the intent or claimed pro-competitive justifications.[68]

In the context of frontier lab collaboration, price fixing concerns may arise if labs agree on pricing structures for safety-enhanced models (for example, two leading labs agree not to offer their most safety-aligned models before a certain price floor) or jointly commit to avoid undercutting each other on price in order to ensure sufficient funding for ongoing risk evaluation and safety research. Even well-intentioned coordination to sustain safety investments could be interpreted as unlawfully limiting price competition and be subject to *per se* treatment.

C      *Market allocation*

Market allocation occurs when competitors agree to divide markets by geography, customer type, or product category, to avoid direct competition. A market allocation agreement is a "classic example" of *per se* illegal conduct that reduces consumer choice and distorts competitive dynamics.[69]

---

[67] See, by way of example, *Arizona v. Maricopa County Medical Society*, 457 U.S. 332, 347 (1982) ("We have not wavered in our enforcement of the per se rule against price fixing"); *United States v. Trenton Potteries Co.*, 273 U.S. 392 (1927).
[68] See *United States v. Aiyer*, 33 F.4th 97 (2d Cir. 2022), holding that it "would have been legal error" to consider claimed procompetitive justifications for price-fixing, "absent a properly asserted exception to the per se rule" (cited in Daniel Francis and Christopher Jon Sprigman, *Antitrust: Principles, Cases, and Materials* (2024) 208).
[69] *United States v. Topco Associates*, 405 U.S. 596, 608 (1972) ("One of the classic examples of a per se violation of § 1 is an agreement between competitors at the same level of the market structure to allocate territories in order to minimize competition").

AI safety collaborations may lead to informal or formal specialisation, where labs implicitly agree to focus on different sectors or capabilities. Even without an explicit agreement, such conduct may be suspect if it reduces competitive overlap. Of course, explicit agreement will also raise antitrust issues — for example, if two leading labs decide that one will focus on enterprise applications, and the other on consumer applications, with an understanding that each will stay out of the other's sector to avoid duplication and inefficiency.

### D  *Sharing competitively sensitive information*

Sharing competitively sensitive information (including future pricing, product strategies, or customer-specific data) can facilitate collusion or coordinated behavior, even in the absence of a formal agreement.

Information sharing between competitors has been described as an "increasingly challenging area for antitrust", in part due to the "speed, volume, and ecosystems of data that firms have access to or collect".[70] AI safety collaborations may require the exchange of technical data, safety research findings, and access to unreleased or non-public models. If these exchanges include commercially sensitive material (for example, projected release

---

[70] Cynthia Hanwalt and Denise Hearn, *Recommendations to Update the FTC & DOJ's Guidelines for Collaborations Among Competitors* (2024).

dates or pricing plans), they risk enabling tacit coordination or softening competitive rivalry.[71]

## V    Policy proposals

The mere perception that antitrust law may prohibit good-faith collaboration could chill participation. While certain conduct like price fixing can readily be avoided through carefully designed collaboration agreements, the potential chilling effect may come from the lack of clarity as to how competitor collaborations would be assessed by antitrust enforcers. This is particularly so in the absence of clear regulatory guidance or safe harbours.[72] This section advocates for the exploration of sensible and carefully calibrated legislative and regulatory measures that would promote essential collaborative safety research.[73] Before surveying possible proposals, it is prudent to make some preliminary observations.

First, it is not clear that legitimate safety collaboration would *actually* violate existing antitrust laws. Most forms of collaboration discussed above — cross red-teaming, sharing incident reports, and joint model development — may be unlikely to raise serious antitrust concerns. If challenged, it is of course possible (and perhaps likely) that most collaborative efforts would be upheld under the evaluative rule of reason analysis, which weighs pro-competitive benefits against anticompetitive harms. However, legal and regulatory uncertainty can act as a powerful deterrent.[74] Frontier AI labs facing intense commercial pressure may conclude that the legal, financial and reputational costs of a regulatory action outweighs the benefits of collaborating for the purposes of working towards safe developmental outcomes.

---

[71] Some difficulty may arise in circumstances where access to non-public models allows a competitor to infer market-sensitive information.

[72] See Helen Toner, Testimony before the U.S. House Committee on the Judiciary, Subcommittee on Courts, Intellectual Property, Artificial Intelligence, and the Internet, "Protecting Our Edge: Trade Secrets and the Global AI Arms Race" (2025) 7–8.

[73] Private governance proposals are outside the scope of this paper. For a useful overview of private governance, see Dean W. Ball, *A Framework for the Private Governance of Frontier Artificial Intelligence* (2025), https://arxiv.org/abs/2504.11501.

Second, antitrust enforcement in practice is not triggered merely by the fact of competitors working together. An antitrust investigation is an incredibly complex, resource-intensive exercise, and the decision to open an investigation is one that requires balancing regulatory priorities and the allocation of scarce public resources. But this may be cold comfort for frontier AI labs collaborating in what may seem like a regulatory "grey area".

Finally, the proposals that follow are not intended to be fully fleshed-out legal instruments. Rather, they are offered as starting points to prompt further discussion among policymakers, regulators, scholars, and industry stakeholders. The purpose of this section is to survey potential proposals for reducing legal uncertainty and both enabling and encouraging collaborative AI safety research.

---

[74] See UTS Human Technology Institute and e61 Institute, *AI, Productivity, and Australia's Choice of Regulatory Framework* (2025) 11.

### A    *Legislative proposals*

Legislation would be the cleanest and most durable method for enabling AI safety collaboration that is protected from unnecessary antitrust scrutiny. However, congressional gridlock in the United States poses a real barrier to new legislative proposals. Notwithstanding, Congress could explore the following options to provide legal clarity and reduce the chilling effect on good-faith cooperation among frontier AI developers. These proposals are designed to facilitate necessary safety collaboration while preserving antitrust enforcement against genuinely anti-competitive conduct.

#### 1    Research and development exemption: expand the National Cooperative Research and Production Act

One possibility to encourage AI safety collaboration is to expand the scope and application of the National Cooperative Research and Production Act of 1993 (NCRPA).[75] The NCRPA was originally enacted to encourage joint R&D efforts, and is a well-

---

[75] 15 U.S.C. §4301.

established mechanism for reducing antitrust exposure where competitors collaborate on research.[76]

At bottom, the NCRPA encourages good-faith collaborations by making clear that the rule of reason should govern collaborations designed to produce a public benefit.[77] It does not provide blanket immunity from antitrust enforcement — rather, it aims to promote innovation and strengthen American competitiveness by:[78]

*Clarifying the applicability of the rule of reason standard to the antitrust analysis of joint ventures and standards development organizations (or "SDOs") while engaged in a standards development activity.*

*Providing for the possible recovery of attorneys fees by joint ventures and SDOs that are prevailing parties in damage actions brought against them under the antitrust laws.*

*Providing to parties to joint ventures and to SDOs the opportunity to limit any possible monetary damages that might be sought from them in actions brought under the antitrust laws to actual — as opposed to treble — damages.*

---

[76] See John T. Scott, *The National Cooperative Research and Production Act*, 2 Issues in Competition Law and Policy 1297 (2008).

[77] See Tom Wheeler and Blair Levin, *With AI, we need both competition and safety* (2024), https://www.brookings.edu/articles/with-ai-we-need-both-competition-and-safety/, citing *United States v. United States Gypsum Co*, 438 U.S. 422 (1978).

[78] Department of Justice, *Filing a Notification under the NCRPA*, https://www.justice.gov/atr/filing-notification-under-ncrpa.

At present, only joint ventures and standards development organisations (SDOs) are entitled to file an NCRPA notification with the DOJ and FTC.[79] A joint venture is defined broadly in the NCRPA as follows:

*The term "joint venture" means any group of activities, including attempting to make, making, or performing a contract can file an NCRPA notification if they engage in the following activities:*

    *(a) theoretical analysis, experimentation, or systematic study of phenomena or observable facts,*

    *(b) the development or testing of basic engineering techniques,*

    *(c) the extension of investigative findings or theory of a scientific or technical nature into practical application for experimental and demonstration purposes, including the experimental production and testing of models, prototypes, equipment, materials, and processes,*

    *(d) the production of a product, process, or service,*

    *(e) the testing in connection with the production of a product, process, or service by such venture,*

    *(f) the collection, exchange, and analysis of research or production information, or*

---

[79] The DOJ provides the requirements to make a notification, the details of which are not necessary to set out in full: Department of Justice, *Filing a Notification under the NCRPA*, https://www.justice.gov/atr/filing-notification-under-ncrpa.

> *(g) any combination of the purposes specified in subparagraphs (A), (B), (C), (D), (E), and (F),*
>
> *and may include the establishment and operation of facilities for the conducting of such venture, the conducting of such venture on a protected and proprietary basis, and the prosecuting of applications for patents and the granting of licenses for the results of such venture, but does not include any activity specified in subsection (b).*

The NCRPA also makes clear that "joint venture" *excludes* certain activities, relevantly including the following activities:

> *(1) exchanging information among competitors relating to costs, sales, profitability, prices, marketing, or distribution of any product, process, or service if such information is not reasonably required to carry out the purpose of such venture,*
>
> *(2) entering into any agreement or engaging in any other conduct restricting, requiring, or otherwise involving the marketing, distribution, or provision by any person who is a party to such venture of any product, process, or service …*
>
> *…*
>
> *(4) entering into any agreement or engaging in any other conduct allocating a market with a competitor …*

These exclusions clearly demonstrate an ongoing admonition of commercially-sensitive information sharing, output restrictions, and market allocation.

The NCRPA definition of "joint venture" appears to be broad enough to include collaborations between competitors that are not enshrined in a contract. But the operative provision of the NCRPA — which makes clear that joint venture activities shall not be deemed per se illegal, but rather analysed under the rule of reason — only applies to "any person in making or performing *a contract* to carry out a joint venture".[80] While frontier AI labs may seek to engage in collaborative safety research by way of a formal contract, this requirement may impose some limitations on exempting desirable collaborations from antitrust scrutiny.

Based on the text of the NCRPA, many potential forms of safety research collaboration among frontier AI labs may be protected under the NCRPA. For example, collaboration between labs to jointly design and test red-teaming protocols, or to exchange technical benchmarks for measuring model robustness, would likely fall within the NCRPA's scope as permissible research or testing activities.[81]

---

[80] 15 U.S.C. § 4302(1) — Rule of reason standard
> In any action under the antitrust laws, or under any State law similar to the antitrust laws, the conduct of—
> (1) any person in making or performing a contract to carry out a joint venture, or
> (2) a standards development organization while engaged in a standards development activity,
> shall not be deemed illegal per se; such conduct shall be judged on the basis of its reasonableness, taking into account all relevant factors affecting competition, including, but not limited to, effects on competition in properly defined, relevant research, development, product, process, and service markets. For the purpose of determining a properly defined, relevant market, worldwide capacity shall be considered to the extent that it may be appropriate in the circumstances.

[81] See 15 U.S.C. § 4301(a)(6)(c): "[I]nvestigative findings or theory of a scientific or technical nature into practical application for experimental and demonstration purposes, including the experimental production and testing of models".

But one of the more important collaborative efforts — development pauses — may not be covered by the NCRPA. This is due to the express exclusion of output restrictions. This could be addressed by narrowly carving out an exception to the output restriction exclusion for safety-driven developmental pauses. Specifically, the NCRPA could be amended to allow output restrictions where they are demonstrably linked to risk mitigation efforts agreed upon through a transparent, time-limited, and reviewable process, such as in response to newly identified model vulnerabilities or emergent capabilities. This exception could be carefully drafted to prevent abuse by limiting its application to specific, narrowly-defined safety triggers.

Just as the original proponents of the NCRPA[82] argued that the "threat of litigation and per se illegality was hindering American technological progress",[83] it appears that those same threats are hindering safe and responsible AI development. With modest amendments to potentially allow for developmental pauses, the NCRPA would be well-suited to address the current chilling effect. It would allow regulators to retain enforcement discretion over truly anticompetitive conduct, while giving labs space to coordinate on essential safety efforts without fear of triggering costly investigations or litigation.

---

[82] At the time of enactment, the National Cooperative Research Act of 1984.
[83] See John T. Scott, *The National Cooperative Research and Production Act*, 2 Issues in Competition Law and Policy 1297, 1300 (2008).

## 2    Permissible information sharing: following the Cybersecurity Information Sharing Act

It is likely that any pragmatically valuable safety collaboration between frontier AI labs will involve some amount of information sharing. Information sharing done in good faith can improve efficiencies and lead to highly beneficial safety outcomes.[84] The Cybersecurity Information Sharing Act of 2015 (CISA) provides an antitrust exemption for information sharing in the context of cybersecurity threats. The CISA exemption is in the following terms:[85]

> *ANTITRUST EXEMPTION.—*
>
> > *(1) IN GENERAL*
> >
> > > *Except as provided in section 1507(e) of this title, it shall not be considered a violation of any provision of antitrust laws for 2 or more private entities to exchange or provide a cyber threat indicator or defensive measure, or assistance relating to the prevention, investigation, or mitigation of a cybersecurity threat, for cybersecurity purposes under this title.*
> >
> > *(2) APPLICABILITY*

---

[84] See generally Department of Justice and Federal Trade Commission, *Antitrust Policy Statement on Sharing of Cybersecurity Information* (2014) 6.
[85] 6 U.S.C. § 1503(e).

> *Paragraph (1) shall apply only to information that is exchanged or assistance provided in order to assist with—*
>
> *(a) facilitating the prevention, investigation, or mitigation of a cybersecurity threat to an information system or information that is stored on, processed by, or transiting an information system; or*
>
> *(b) communicating or disclosing a cyber threat indicator to help prevent, investigate, or mitigate the effect of a cybersecurity threat to an information system or information that is stored on, processed by, or transiting an information system.*

(Relevant terms are defined elsewhere in the CISA).

This CISA exemption could act as a model for an AI safety-related antitrust exemption. Indeed, the Law Reform Institute is currently working on a draft piece of legislation titled "Collaboration on Frontier Model Risks Act", which would include the following provision:[86]

> *SEC. 2. ANTITRUST EXEMPTION*

---

[86] Law Reform Institute, *Antitrust Exemption for AI Frontier Model Risks* (Draft at 30 May 2025). The Law Reform Institute also proposes a carveout to allow the DOJ and FTC to seek an injunction to "prohibit the initiation or continuation of any anticompetitive conduct that, but for section 2, would have violated antitrust laws".

> *Except as provided in section 3, it shall not be considered a violation of any provision of <u>antitrust laws</u> to provide or exchange information or <u>assistance</u> relating to a <u>frontier model risk</u> for the purpose of preventing, investigating, or mitigating a <u>frontier model risk</u>.*

In my view, this proposal has considerable merit. It should serve the intended purpose of alleviating concerns among frontier AI labs that good-faith information sharing for the purposes of risk-mitigation could lead to unwanted antitrust scrutiny. Like the cybersecurity context, many frontier model risks are high-impact threats that would benefit from real-time information sharing. By adopting a similar exemption, policymakers would be drawing on an established legislative precedent that offers frontier AI labs with the certainty and clarity needed to act decisively in the public interest.

### B    *Regulatory guidance*

Regulatory guidance offers a more flexible and immediately actionable alternative to legislative reform. The DOJ and FTC can issue clarifying statements or policy guidance to signal how they will evaluate collaborative safety efforts under existing antitrust law. Formal guidance can play a valuable role in reducing legal uncertainty and encouraging collaborative AI safety research. Although it is prudent to note that guidance alone may not fully insulate labs from legal risk — private plaintiffs can still pursue antitrust cases, and shifts in administration may lead to changes in enforcement priorities.[87]

**1      Reinstate collaboration guidelines and create an AI safe harbour**

In 2000, the DOJ and FTC issued "Antitrust Guidelines for Collaborations Among Competitors" (Guidelines). In December 2024, the agencies agreed to withdraw the Guidelines. Given the subsequent changes in leadership at each agency, and the fact that the two dissenting members of the FTC now make up the ideological majority, there may be appetite to draft new Guidelines. The agencies should consider introducing a new set of Guidelines that include an express safe harbour for collaborative AI safety research and development, or guidance that such collaborations would be assessed under the rule of reason.

The Guidelines were designed to "encourag[e] procompetitive collaborations [and] deter[] collaborations likely to harm competitions and consumers".[88] To that end, the Guidelines provided the analytical framework that the FTC and DOJ would use to evaluate the competitive effects of collaborations between competitors. For example, it instructed parties that while courts may conclusively presume certain agreements are *per se* illegal, in circumstances where "participants in an efficiency-enhancing integration of economic activity enter into an agreement that is reasonably related to the integration and reasonably necessary to achieve its procompetitive benefits, the Agencies analyze the agreement under the rule of reason, even if it is of a type that might otherwise be considered per se illegal". The Guidelines did not bind antitrust enforcers, and each agency would still "evaluate each case in light of its own facts", but they would "apply the analytical framework set forth in [the] Guidelines reasonably and flexibly".[89]

---

[87] There can also be shifts in the enforcement priorities within the same administration.
[88] Department of Justice and Federal Trade Commission, *Guidelines* at 2.

A key feature of the Guidelines was the creation of antitrust "safety zones" to provide certainty for procompetitive collaborations.[90] These safety zones were designed to encourage collaborations that were so unlikely to have anticompetitive effects that they would be presumed lawful without an inquiry into their specific circumstances. The safety zones applied where there existed countervailing competitive pressures on the collaborating parties (for example: where the combined market share of collaborators was 20% or less in a relevant market;[91] or, in the case of research and development, where at least three other "independently controlled research efforts" could serve as close substitutes to the collaboration's research and development activities.)[92] While the market share and the competitive research efforts limitations would likely not prove helpful in the case of frontier AI lab collaboration, they serve as a useful model for the agencies' flexibility and regulatory discretion. If the agencies are attuned to the benefit of collaborative AI safety research (and the risks of deterring it), it may create an appetite to issue regulatory guidance that removes the perceived antitrust barrier to collaboration.

---

[89] Department of Justice and Federal Trade Commission, *Guidelines* at 2.
[90] Department of Justice and Federal Trade Commission, *Guidelines* at 24.
[91] Department of Justice and Federal Trade Commission, *Guidelines* at 25.
[92] Department of Justice and Federal Trade Commission, *Guidelines* at 26–27.

Reviving and updating these Guidelines could provide a clear framework for AI labs to collaborate on safety research. Any new guidelines should offer a clear indication that certain collaborations either fall within a safe harbour, or will be presumed to be subject to rule of reason analysis.[93] By providing clear rules, the agencies can encourage safety-oriented collaborations while still deterring truly anticompetitive conduct. Alternatively, the DOJ and FTC could consider releasing a joint policy statement to a similar effect, whereby the agencies take a position that certain forms of collaboration should not raise antitrust concerns (as was done in the cybersecurity context).[94] Regardless of whether regulatory guidance comes in the form of detailed guidelines or a policy statement, it would serve an important expressive function — a clear signal from antitrust enforcers that collaboration for the purposes of safety and risk reduction is encouraged.[95]

---

[93] Additionally, revised guidelines could provide for information sharing. See Cynthia Hanwalt and Denise Hearn, *Recommendations to Update the FTC & DOJ's Guidelines for Collaborations Among Competitors* (2024) 25–26.

[94] Department of Justice and Federal Trade Commission, *Antitrust Policy Statement on Sharing of Cybersecurity Information* (2014).

[95] See generally Cass R. Sunstein, *On the Expressive Function of Law*, U. Penn. L.R. (1996).

## 2   Business letter review

The DOJ provides for a "business review procedure", which allows businesses to "determine how the [Antitrust] Division may respond to proposed joint ventures or other business conduct".[96] Frontier AI labs could make a business letter request to seek guidance from the DOJ by way of analysis and commentary on the possible competitive impact of a proposed collaboration. Business review requests usually concern joint ventures or information sharing (although other proposed courses of conduct can be the subject of a request).[97] The DOJ instructs that it will generally provide one of three responses to a business review request:[98]

> "The Department of Justice does not presently intend to bring an enforcement action against the proposed conduct.
>
> The DOJ declines to state its enforcement intentions. The Division may or may not file suit if the proposed conduct happens.
>
> The Department of Justice will sue if the proposed conduct happens."

This is merely a statement of present intent, meaning the DOJ is not prevented from taking subsequent action.[99] At present, there is seemingly no legal barrier to frontier AI labs requesting a business review. One possible reason why this has not yet happened is the risk that the DOJ declines to state its intentions and creates further uncertainty and the fear of investigation. If the DOJ believes it is desirable to encourage safety collaborations, it could proactively encourage frontier AI labs to seek a business review. This would likely provide some level of comfort to engage with antitrust enforcers to work towards safe developmental outcomes.

---

[96] 28 C.F.R. § 50.6. See also Department of Justice, *Introduction to Antitrust Division Business Reviews* (2011).
[97] See Department of Justice, *What Is a Business Review?*, https://www.justice.gov/atr/what-business-review.
[98] See Department of Justice, *Introduction to Antitrust Division Business Reviews*, https://www.justice.gov/sites/default/files/atr/legacy/2011/11/03/276833.pdf.
[99] See *United States v. Grinnell Corporation*, 30 F.R.D. 358, 363 ("[A] "present intention not to take action" [cannot] be equated with future immunity.")

## VI    Conclusion

As frontier AI labs race to develop increasingly powerful systems, there is an ever-increasing need to ensure safe and responsible development. Collaboration amongst frontier AI labs can help mitigate developmental risks and prevent catastrophe. While many forms of collaboration may survive antitrust scrutiny, uncertainty can chill such safety-oriented efforts. Frontier AI labs are avoiding deep, substantive engagement with competitors that could ensure safer timelines out of fear of legal, financial, and reputational exposure.

This paper has proposed several sensible legislative and regulatory proposals that could provide much-needed certainty to encourage collaboration. But these proposals are not exhaustive or final. They are a starting point to stir discussion, debate, and further analysis. Future work could examine other domestic tools (such as § 708 of the Defense

Production Act)[100] and international mechanisms (such as the Australian Competition and Consumer Commission's class exemption procedure or the European Commission's horizontal block exemption regulations) as avenues for safeguarding essential safety collaboration from antitrust scrutiny. Another issue is that of regulatory design and how pro-safety measures can be constructed in such a way to mitigate against vulnerabilities that might otherwise be exposed by encouraging coordination.

Competition policy has laudable goals — protecting consumers, preventing misuses of market power, and encouraging responsible innovation. But the real or perceived strictures of U.S. antitrust law should not unduly restrict collaborations that are designed and executed, in good faith, to prevent catastrophic AI risks. Striking the right balance between promoting competition and enabling responsible coordinated safety research will require careful analysis and sustained engagement between industry and policymakers.

---

[100] Section 708 empowers the President to "consult with representatives of industry … in order to provide for the making by such persons, with approval of the President, of voluntary agreements and plans of action to help provide for the national defense". It also establishes a special defense that may shield companies cooperating under the Defense Production Act from antitrust liability "with respect to any action taken to develop or carry out any voluntary agreement or plan of action".